\documentclass[useAMS,usenatbib]{mn2e}
\usepackage{graphicx,savefnmark}
\usepackage{float}
\usepackage{color}
\usepackage{booktabs}
\usepackage{threeparttable}
\usepackage{supertabular,amsmath}

\title[Radio Afterglows and Host Galaxies of GRBs]{Radio Afterglows and Host Galaxies of Gamma-Ray Bursts}
\author[]
         {Long-Biao Li$^{1}$, Zhi-Bin Zhang$^{1,9}$\thanks{E-mail: sci.zbzhang@gzu.edu.cn},
         Yong-Feng Huang$^{2}$\thanks{E-mail: hyf@nju.edu.cn}, Xue-Feng Wu$^{3}$,
         \and Si-Wei Kong$^{4}$, Di Li$^{5,6}$, Heon-Young Chang$^{7}$ and Chul-Sung Choi$^{8}$ \\
$^{1}$ Guizhou University, Department of Physics, College of Sciences, Guiyang 550025, China\\
$^{2}$ Department of Astronomy, Nanjing University, Nanjing 210093, China\\
$^{3}$ Purple Mountain Observatory, Chinese Academy of Sciences, Nanjing 210008, China\\
$^{4}$ Institute of High Energy Physics, Chinese Academy of Sciences, 19B Yuquan Road,
       Beijing 100049, China\\
$^{5}$ National Astronomical Observatories of China, Chinese Academy of Sciences,
       20A Datun Road, Beijing 100020, China\\
$^{6}$ Key Laboratory of Radio Astronomy, Chinese Academy of Sciences, China\\
$^{7}$ Department of Astronomy and Atmospheric Sciences, Kyungpook National University,
       1370 Sankyuk-dong, Buk-gu, \\Daegu 702-701, Republic of Korea\\
$^{8}$ Korea Astronomy and Space Science Institute, 36-1 Hwaam, Yusong, Daejon 305-348,
       Republic of Korea\\
$^{9}$ Department of Physics and Astronomy, University of Nevada, Las Vegas, NV 89012, USA\\}

\begin{document}

\pagerange{\pageref{firstpage}--\pageref{lastpage}} \pubyear{2014}

\maketitle

\label{firstpage}

\begin{abstract}
  Considering the contribution of the emission from the host galaxies of gamma-ray bursts (GRBs) to the radio afterglows,
  we investigate the effect of host galaxies on observations statistically. For the three types of events, e.g.
  low-luminosity, standard and high-luminosity GRBs, it is found that a tight correlation exists between
  the ratio of the radio flux (RRF) of host galaxy to the total radio peak emission and the observational frequency.
  Especially, toward lower frequencies, the contribution from the host increases significantly.
  The correlation can be used to get a useful estimate for the radio brightness of those host galaxies which
  only have very limited radio afterglow data. Using this prediction, we re-considered the theoretical radio afterglow
  light curves for four kinds of events, i.e. high-luminosity, low-luminosity, standard and failed GRBs,
  taking into account the contribution from the host galaxies and aiming at exploring the detectability of these events
  by the Five-hundred-meter Aperture Spherical radio Telescope (FAST). Lying at a typical redshift of $z=1$,
  most of the events can be detected by FAST easily. For the less fierce low-luminosity GRBs,
  their radio afterglows are not strong enough to exceed the sensitivity limit of FAST at such distances.
  However, since a large number of low luminosity bursts actually happen very near to us,
  it is expected that FAST will still be able to detect many of them.
\end{abstract}


\section{Introduction}

\begin{table*}
  \caption{Observational properties of \textbf{50} GRBs and their host galaxies in radio bands.}
  \scriptsize
  \centering
  \begin{minipage}{172mm}
  \begin{threeparttable}
  \begin{tabular}{@{}lccccccccc@{}}
  \hline
  GRB   & $E_{iso}$\tnote{a} & $T_{90}$ \tnote{a}  & z \tnote{a}  &  Frequency   & Host Flux Density
              & Peak Flux Density \tnote{b} & Peak Time    & References\tnote{c}   & RRF\\

        & ($10^{51}$ erg)    & (s)                 &              &  (GHz)       & ($\mu$Jy)
              &  ($\mu$Jy)         & (days)           &                      & ($\%$)  \\

    (1) &  (2)   & (3)  & (4) &   (5)         &   (6)              & (7) &   (8)    & (9)   & (10)   \\

  \hline
      \multicolumn{10}{|c|}{Low-Luminosity GRBs}   \\
  \hline
  020903\tnote{n}
         & 0.023  & 13 & 0.250  & 1.43 &  $43\pm66$    & $294\pm91$    & $36.73$       & 28,9  & $14.6\pm14.8$ \\
         &        &    &        & 4.86 &  $21\pm34$    & $782\pm28$    & $36.7\pm1.3$  & 28,9  & $2.7\pm4.4$ \\
         &        &    &        & 8.46 &  $190\pm30$   & $1058\pm19$   & $23.80$       & 28,9  & $18.0\pm2.9$  \\
  031203\tnote{n}
         & 0.115  & 30 & 0.105  & 1.43 &  $254\pm46$\tnote{h}
                                                       & $929\pm60$    & $65.5\pm5.3$  & 22,9  & $27.3\pm5.3$  \\
         &        &    &        & 4.86 &  $393\pm60$   & $828\pm28 $   & $58.4\pm2.1$  & 27,9  & $47.5\pm7.4$  \\
         &        &    &        & 8.46 &  $53\pm52$    & $724\pm19 $   & $48\pm1.3$    & 27,9  & $7.3\pm7.2$   \\
  050416A\tnote{n}
         & 1.00   & 3  & 0.650  & 4.86 & $50.5\pm41.8$ & $485\pm36$    & $48.5\pm3.5$  & 30,9  & $10.4\pm8.7$  \\
         &        &    &        & 8.46 &  $20\pm51$    & $373\pm36$    & $49\pm3.6$    & 30,9  & $7.3\pm13.7$  \\

  060218\tnote{n}
         & 0.003  & 128& 0.033  & 1.43 & $69.5\pm69.5$ &  $198\pm58$   & $4.9$         & 29,9  & $35.1\pm36.6$ \\
         &        &    &        & 4.86 &  $32\pm32$    &  $245\pm50$   & $3.8\pm0.7$   & 29,9  & $13.1\pm13.3$ \\
         &        &    &        & 8.46 & $22.8\pm20.5$ &  $471\pm83$   & $2\pm0.3$     & 29,9  & $4.8\pm4.4$   \\
  100418A\tnote{n}
         & 0.520  & 7  & 0.620  & 8.46 & $366\pm53$    &  $1218\pm12$  & $47.6\pm0.6$  & 23,9  & $30.1\pm4.4$  \\
  \hline
      \multicolumn{10}{|c|}{Standard GRBs}   \\
  \hline
  970508 & 7.10   & 14 & 0.835  & 1.43 & $100\pm32$    &  $381\pm19$   & $179.1\pm7.6$ & 15,9  & $26.3\pm8.5$  \\
         &        &    &        & 4.86 & $104\pm44$    &  $780\pm13$   & $57.6\pm0.9$  & 15,9  & $13.3\pm5.6$  \\
         &        &    &        & 8.46 & $8\pm19$      &  $958\pm11$   & $37.2\pm0.4$  & 17,9  & $0.8\pm2.0$   \\
  980703 & 69.0   & 90 & 0.966  & 1.43 & $68\pm6.6$\tnote{h}
                                                       &  $263\pm81$   & $25.4\pm5.5$  &  3,9  & $25.9\pm8.4$  \\
         &        &    &        & 4.86 & $42.1\pm8.6$\tnote{h}
                                                       &  $1055\pm30$  & $9.1\pm0.2$   &  3,9  & $4.0\pm0.8$   \\
         &        &    &        & 8.46 & $39.3\pm4.9$\tnote{h}
                                                       &  $1370\pm30$  & $10\pm0.2$    &  3,9  & $2.9\pm0.4$   \\
  981226 & 5.90   & 20 & 1.11   & 8.46 & $12\pm22.6$   &  $137\pm34$   & $8.2\pm1.6$   & 12,9  & $8.8\pm1.7$   \\
  000301C& 43.7   & 10 & 2.034  & 4.86 & $85\pm33$     &  $240\pm53$   & $4.3$         &  1,9  & $35.4\pm15.8$ \\
         &        &    &        & 8.46 & $18\pm7$\tnote{h}
                                                       &  $520\pm24$   & $14.1\pm0.5$  & 24,9  & $3.5\pm1.4$   \\
  000418 & 75.1   & 30 & 1.119  & 1.43 & $48\pm15$\tnote{h}
                                                       & $210\pm180$\tnote{$\dagger$}
                                                                       &  $16.56$      &  4,2  & $22.9\pm20.9$ \\
         &        &    &        & 4.86 & $110\pm52$    & $897\pm39$    & $27\pm1$      &  4,9  & $12.3\pm5.8$ \\
         &        &    &        & 8.46 & $37\pm12$\tnote{h}&$1085\pm22$& $18.1\pm0.3$  &  4,9  & $3.4\pm1.1$   \\
  010921 & 9.00   & 24 & 0.450  & 4.86 & $27\pm25$     &  $161\pm20$   & $31.5\pm3.3$  & 17,9  & $16.8\pm15.7$ \\
         &        &    &        & 8.46 & $43.7\pm21.3$ &  $229\pm22$   & $27.01$       & 17,9  & $19.1\pm9.5$  \\
  011211 & 63.0   & 400& 2.140  & 8.46 & $24\pm16$     &  $162\pm13$   & $13.2\pm1.6$  & 17,9  & $14.8\pm9.9$  \\
  020819B& 7.90   & 50 & 0.410  & 8.46 & $44.7\pm23$   &  $291\pm21$   & $12.2\pm1.1$  & 20,9  & $15.4\pm8.0$  \\
  021004 & 38.0   & 50 & 2.330  & 4.86 & $78\pm28$     &  $470\pm26$   & $32.2\pm1.4$  & 17,9  & $16.6\pm6.0$  \\
         &        &    &        & 8.46 & $59.4\pm19.5$ &  $780\pm23$   & $18.7\pm0.5$  & 17,9  & $7.6\pm2.5$   \\
  021211\tnote{n}
         & 11.0   & 8  & 1.010  & 8.46 & $12\pm36$     &  $60\pm28$\tnote{$\dagger$}
                                                                       &  $8.85$       & 11,11  & $20.0\pm60.7$ \\
  030329\tnote{n}
         & 18.0   & 63 & 0.169  & 1.43 & $450\pm109$   &  $2232\pm30$  & $78.6\pm2.1$  &  5,9  & $20.3\pm4.9$  \\
         &        &    &        & 4.86 & $332\pm51.7$  &  $10337\pm33$ & $32.9\pm0.1$  &  5,9  & $3.2\pm0.5$   \\
         &        &    &        & 8.46 & $248\pm48.2$  &  $19567\pm28$ & $17.3\pm0.1$  &  5,9  & $1.3\pm0.3$   \\
  050824\tnote{n}5
         & 1.50   & 23 & 0.830  & 8.46 & $36\pm27$     &  $156\pm34$   & $7.2$         & 17,9  & $23.1\pm18.0$ \\
  070612A& 9.12   & 369& 0.617  & 4.86 & $222\pm42.3$  &  $580\pm20$   & $140.3\pm5.8$ & 17,9  & $38.3\pm7.4$  \\
         &        &    &        & 8.46 & $68\pm49$     &  $1028\pm16$  & $84.1\pm2$    & 17,9  & $6.6\pm4.8$   \\
  071010B& 26.0   & 36 & 0.947  & 4.86 & $57\pm60.4$   &  $227\pm114$  & $12.5\pm4.4$  & 17,9  & $25.1\pm29.5$ \\
         &        &    &        & 8.46 & $17\pm47$	   &  $341\pm41$   & $4.2\pm0.5$   & 17,9  & $5.0\pm13.8$  \\
  090328 & 100    & 57 & 0.736  & 8.46 & $21\pm64$     &  $686\pm26$   & $16.1\pm0.7$  &  7,9  & $3.1\pm9.3$   \\
  091020 & 45.6   & 39 & 1.710  & 8.46 & $90\pm41.1$   &  $399\pm21$   & $10.9\pm0.6$  & 17,9  & $22.6\pm10.4$ \\
  100814A& 59.7   & 175& 1.44   & 4.50 & $85\pm27$     &  $496\pm24$   & $13\pm1$      & 17,9  & $17.1\pm5.5$  \\
  100901A& 17.8   & 439& 1.408  & 4.50 & $75\pm40$     &  $331\pm30$   & $4.9$         & 17,9  & $22.7\pm12.3$ \\
  \hline
      \multicolumn{10}{|c|}{High-Luminosity GRBs}   \\
  \hline
  970828 & 296    & 147& 0.958  & 1.43 & $14\pm22$     &  $57\pm51$\tnote{$\dagger$}
                                                                       & $208.1$       & 10,10  & $24.6\pm44.4$ \\
         &        &    &        & 8.46 & $32\pm20.5$   &  $144\pm31$   & $7.8\pm1.8$   & 10,9   & $22.2\pm15.0$ \\
  980329 & 2100   & 58 & 2--3.9 & 1.43 & $38\pm25.5$   &  $134\pm43$\tnote{$\dagger$}
                                                                       & $113.54$      & 32,32  & $28.4\pm21.1$ \\
         &        &    &        & 4.86 & $22\pm37$     &  $171\pm14$   & $91.5\pm11.3$ & 32,9   & $12.9\pm21.7$ \\
         &        &    &        & 8.46 & $13\pm11$     &  $332\pm11$   & $33.5\pm1.4$  & 32,9   & $3.9\pm3.3$   \\
  990123 & 2390   & 100& 1.600  & 8.46 & $7.7\pm11.1$  &  $260\pm32$\tnote{$\dagger$}
                                                                       & $24.65$       & 21,21  & $3.0\pm4.3$   \\
  990506 & 949    & 220& 1.307  & 8.46 & $31\pm27.6$   &  $581\pm45$\tnote{$\dagger$}
                                                                       & $2.70$        & 31,31  & $5.3\pm4.8$   \\
  991208\tnote{n}
         & 110    & 60 & 0.706  & 1.43 & $54.8\pm44.4$ &  $263\pm49$   & $8.9\pm1.8$   & 18,9   & $20.8\pm17.3$  \\
         &        &    &        & 8.46 & $37.3\pm25.8$ &  $1804\pm24$  & $7.8\pm0.1$   & 18,9   & $2.1\pm1.4$   \\
  991216 & 675    & 25 & 1.020  & 8.46 & $20\pm24.7$   &  $960\pm67$\tnote{$\dagger$}
                                                                       & $1.49$        & 13,13  & $2.1\pm2.6$   \\
  000210 & 200    & 10 & 0.850  & 8.46 & $18\pm9$\tnote{h}
                                                       &  $93\pm21$\tnote{$\dagger$}
                                                                       & $8.59$        &  4,25  & $19.4\pm10.6$ \\
  000911\tnote{n}
         & 880    & 500& 1.059  & 4.86 & $10\pm25$     &  $71\pm23$    & $11.1$        & 26,9   & $14.1\pm35.5$ \\
         &        &    &        & 8.46 & $33.7\pm25.2$ &  $263\pm33$   & $3.1\pm0.3$   & 26,9   & $12.8\pm9.7$  \\
  000926 & 270    & 25 & 2.039  & 4.86 & $58\pm41.5$   &  $460\pm31$   & $16.9\pm1.5$  & 19,9   & $12.6\pm9.1$  \\
         &        &    &        & 8.46 & $23\pm9$\tnote{h}
                                                       & $629\pm24$    & $12.1\pm0.5$  & 24,9   & $3.7\pm1.4$   \\
  010222 & 133    & 170& 1.477  & 4.86 & $23\pm8$\tnote{h}& $144\pm47$ & $2.3$         & 24,9   & $16.0\pm7.6$  \\
         &        &    &        & 8.46 & $38.8\pm19.8$ &  $344\pm39$\tnote{$\dagger$}
                                                                       &$1.35$         & 17,17  & $11.3\pm5.9$  \\
  020813 & 800    & 113& 1.254  & 4.86 & $23\pm42$     &  $349\pm43$\tnote{$\dagger$}
                                                                       & $5.15$        & 17,17  & $6.6\pm12.1$  \\
         &        &    &        & 8.46 & $30.8\pm19.8$ &  $323\pm39$\tnote{$\dagger$}
                                                                       & $1.24$        & 17,17  & $9.5\pm6.2$   \\
  030226 & 120    & 69 & 1.986  & 8.46 & $35\pm18$     &  $171\pm23$   & $6.7\pm1$     & 17,17  & $20.5\pm10.9$ \\
  050603 & 500    & 12 & 2.821  & 8.46 & $79\pm49$     &  $377\pm53$   & $14.1\pm1.8$  & 17,17  & $20.9\pm13.3$ \\
  050820A& 200    & 240& 2.615  & 4.86 & $21\pm51$     &  $256\pm78$\tnote{$\dagger$}
                                                                       & $2.15$        & 17,6   & $8.2\pm20.1$  \\
         &        &    &        & 8.46 & $76\pm30$     &  $634\pm62$\tnote{$\dagger$}
                                                                       & $0.93$        &  6,6   & $12.0\pm4.9$  \\
  050904 & 1300   & 174& 6.290  & 8.46 & $13\pm27$     &  $76\pm14$    & $35.3\pm$1.5  & 16,9   & $17.1\pm35.7$ \\
  051022 & 630    & 200& 0.809  & 8.46 & $41\pm23$     &  $268\pm32$   & $5.2\pm0.7$   & 17,9   & $15.3\pm8.8$  \\
  \hline
  \hline
  \end{tabular}
  \end{threeparttable}
  \end{minipage}
\end{table*}

\begin{table*}
  \renewcommand \thetable {1 (continued)}
  \caption{Observational properties of 50 GRBs and their host galaxies in radio bands.}
  \scriptsize
  \centering
  \begin{minipage}{172mm}
  \begin{threeparttable}
  \begin{tabular}{@{}lccccccccc@{}}
  \hline
  \hline
  GRB   & $E_{iso}$\tnote{a} & $T_{90}$ \tnote{a}  & z \tnote{a}  &  Frequency   & Host Flux Density
              & Peak Flux Density \tnote{b} & Peak Time  & References\tnote{c}   & RRF \\

        & ($10^{51}$ erg)    & (s)                 &              &  (GHz)       & ($\mu$Jy)
              &  ($\mu$Jy)             &     (days)      &            &   ($\%$)  \\

    (1) &  (2)   & (3)  & (4) &   (5)         &   (6)              & (7) &   (8)    & (9)   & (10)   \\
  \hline
                \multicolumn{10}{|c|}{High Luminosity GRBs}   \\
  \hline
  070125 & 955    & 60 & 1.548  & 4.86 & $133\pm21$    &  $308\pm78$   & $27.9$        &  8,9   & $43.2\pm12.9$ \\
         &        &    &        & 8.46 & $64\pm18$     &  $1028\pm16$  & $84.1\pm2$    &  8,9   & $6.2\pm1.7$   \\
  071003 & 324    & 148& 1.604  & 4.86 & $93\pm52$     &  $224\pm54$   & $3.8$         & 17,9   & $41.5\pm25.3$  \\
         &        &    &        & 8.46 & $109\pm45$    &  $616\pm57$   & $6.5\pm0.5$   & 17,9   & $17.7\pm7.5$  \\
  090323 & 4100   & 133& 3.57   & 8.46 & $27\pm38$     &  $243\pm13$   & $15.6\pm1$    &  7,9   & $11.1\pm15.7$ \\
  090902B& 3090   & ...& 1.883  & 8.46 & $18\pm16$     &  $84\pm16$    & $14.1\pm2.6$  &  7,9   & $21.4\pm19.5$  \\
  100414A& 779    & 26 & 1.368  & 8.46 & $20\pm16$     &  $524\pm19$   & $8\pm0.3$     & 17,9   & $3.8\pm3.1$   \\
  \hline
  \multicolumn{10}{|c|}{GRBs without known redshifts} \\
  \hline
  980519 & ...    & 30 & ...    & 4.86 & $25\pm27$     &  $330\pm47$   & $17.9\pm2.4$  & 14,9  & $7.6\pm8.3$   \\
         &        &    &        & 8.46 & $49\pm28$     &  $205\pm23$   & $12.6\pm1.3$  & 14,9  & $23.9\pm13.9$ \\
  001007 & ...    & 375& ...    & 8.46 & $34\pm61$     &  $222\pm33$   & $7.1$         & 17,9  & $15.3\pm27.6$ \\
  001018 & ...    & 31 & ...    & 8.46 & $70\pm23.3$   &  $590\pm68$   & $4.7\pm0.6$   & 17,9  & $11.9\pm4.2$  \\
  021206 & ...    & 20 & ...    & 4.86 & $58\pm24$     &  $480\pm69$   & $7\pm0.8$     & 17,9  & $12.1\pm5.3$  \\
  041219A& ...    & 6  & ...    & 4.86 &  $33\pm48$    &  $473\pm28$   & $6.3\pm0.9$   & 17,9  & $7.0\pm10.2$  \\
         &        &    &        & 8.46 &  $63\pm42$    &  $518\pm150$\tnote{$\dagger$}
                                                                       & $1.69$        & 17,17  & $12.2\pm8.8$   \\
  050713B& ...    & 125& ...    & 4.86 & $44\pm69$     &  $623\pm53$\tnote{$\dagger$}
                                                                       & $21.01$       & 17,17   & $7.1\pm11.1$ \\
         &        &    &        & 8.46 & $59\pm40.5$   &  $373\pm24$   & $14.1\pm0.9$  & 17,9   & $15.8\pm10.9$ \\
  \hline
  \\
  \end{tabular}
  \textbf{Notes.}
    \begin{tablenotes}
       \scriptsize
       \item[a] Refer to Chandra \& Frail (2012).
       \item[b] Observed radio peak flux density.
       \item[c] References are in the following order: host radio flux density, radio peak flux density.\\
                Abbreviations for the references are as follows: (1) Berger et al. (2000), (2) Berger et al. (2001a),
          (3) Berger, Kullarni \& Frail (2001b), (4) Berger et al. (2003a), (5) Berger et al. (2003c),
          (6) Cenko et al. (2006), (7) Cenko et al. (2011), (8) Chandra et al. (2008), (9) Chandra \& Frail (2012),
          (10) Djorgovski et al. (2001), (11) Fox et al. (2003), (12) Frail et al. (1999), (13) Frail et al. (2000a),
          (14) Frail et al. (2000b), (15) Frail, Waxman \& Kulkarni (2000), (16) Frail et al. (2006),
          (17) Frail \& Chandra (2014, private communication), (18) Galama et al. (2003), (19) Harrison et al. (2001),
          (20) Jakobsson et al. (2005), (21) Kulkarni et al. (1999), (22) Micha{\l}owski et al. (2012), (23) Moin et al. (2013),
          (24) Perley \& Perley (2013), (25) Piro et al. (2002), (26) Price et al. (2002), (27) Soderberg et al. (2004a),
          (28) Soderberg et al. (2004b), (29) Soderberg et al. (2006), (30) Soderberg et al. (2007),
          (31) Taylor et al. (2000), (32) Yost et al. (2002).
       \item[h] Host flux densities have been reported in Berger, Kullarni \& Frail (2001b), Berger et al. (2003a),
                Micha{\l}owski et al. (2012) and Perley \& Perley (2013).
       \item[n] SN/GRB, i.e. SN-associated GRBs.
       \item[$\dagger$] These values are the maximum observed flux densities, which are taken as the observed peak flux densities.\\
    \end{tablenotes}
  \end{threeparttable}
  \end{minipage}
\end{table*}

Gamma-ray bursts (GRBs) are bright flashes of gamma-rays that happen randomly in the sky. They were serendipitously
discovered in 1967 by the Vela satellites (Klebesadel et al. 1973), but were poorly understood until Feb 28, 1997
when the first afterglow was detected (Groot et al. 1998). Subsequently, the counterpart of GRB 970508 was detected
as the first afterglow in radio bands (Frail et al. 1997). The detection of afterglows makes it possible for us to
obtain broadband observational data, to identify the host galaxies, and to determine the redshifts of GRBs.
The so called fireball-shock model was developed to explain the main features of GRBs and their afterglows
(Rees \& M\'{e}sz\'{a}ros 1994; Piran 1999; Zhang 2007),
the latter are generally believed to arise from the interaction of the fireball with the surrounding interstellar
medium (ISM) (M\'{e}sz\'{a}ros \& Rees 1997; Prian 2000; M\'{e}sz\'{a}ros 2002).

According to the fireball-shock model (e.g. M\'{e}sz\'{a}ros 2002; Piran 2004; Zhang \& M\'{e}sz\'{a}ros 2004; Zhang 2014),
the outflow of a GRB interacts with the ISM to form an external shock. The shock accelerates electrons.
At the same time, a fraction of the shock energy is transferred to the magnetic field.
The afterglow emission arises from synchrotron radiation of these accelerated electrons due
to their interaction with the magnetic filed. Within this framework, the main features of
GRB afterglows can be well explained.

With more and more afterglows being detected, the study of GRBs has entered an era of full wavelengths
(e.g. Gehrels \& Razzaque 2013). However, according to Hancock, Gaensler \& Murphy (2013),the detection rate
of afterglows is only $\sim 30$ \% at radio wavelengths, although some authors have recently compiled several
larger datasets (de Ugarte Postigo et al. 2012; Chandra \& Frail 2012; Ghirlanda et al. 2013; Staley et al. 2013).
Note that the search for radio afterglows is already in great depth (Chandra \& Frail 2011). Another exclusive
property of radio afterglows is their detection at high redshifts (Frail et al. 2006; Chandra et al. 2010).
For example, there are about 350 GRBs with redshifts measured, of which $\sim$ 32\% are at redshifts $z>2$,
$\sim$ 7\% at $z>4$ and $\sim$ 2\% at $z>6$. The maximum redshift is 9.4 for GRB 090429B, indicating that
GRBs are potentially powerful probes of the early Universe. In fact, GRBs' high luminosities make them potentially
detectable up to very high redshifts (Lamb \& Reichart 2000). They may be observable with the current VLA
up to $z\sim30$ at frequencies higher than $\sim 5$ GHz (Ioka \& M\'{e}sz\'{a}ros 2005; Zhang et al. 2015; Mesler et al. 2014).

China's Five-hundred-meter Aperture Spherical radio Telescope (FAST, Nan et al. 2011) is expected to be
completed in Sep. 2016 and will be the largest radio telescope in the world by then. FAST's receivers will
cover both low (70 MHz-0.5 GHz) and middle (0.5-3 GHz) frequency ranges. Compared with the Arecibo radio
telescope, FAST is expected to be three times more sensitive, with the tracking speed 10 times higher.
Zhang et al. (2015, hereafter Paper I) estimated the exact sensitivity of FAST to GRB afterglows.
They calculated the light curves of radio afterglows of typical low-luminosity GRBs, high-luminosity
GRBs, failed GRBs and standard GRBs in different observational bands of FAST, and found that almost
all types of radio afterglows, except those from low-luminosity GRBs, could be detected by FAST up to $z=5$.
They also argued that FAST can even detect radio afterglows at $\nu=2.50$ GHz up to $z=15$ or even more.
However, they did not consider the contribution from the host galaxies of GRBs. In fact, the observed radio
emission should include the afterglow component as well as the contribution from its host galaxy. In order to
model the observed radio afterglows and evaluate the detectability of FAST more realistically, it is necessary
to consider the contribution from the host galaxy.

In this study, we have collected a large sample of radio afterglows. We examine the relative
contribution from the afterglows and their hosts based on our sample. For this purpose,
we propose a useful method based on the flux ratio of the host to the afterglow to estimate radio
flux density at a given frequency for different kinds of bursts. Taking into account the
contribution from the hosts, we then re-examine the detectability of FAST for those weak radio
afterglows at very high redshifts. Our paper is organized as follows. We briefly introduce
the GRBs of our sample in Section 2. In Section 3, we present our statistic results on the
ratio of the host flux to radio peak flux (RRF). In Section 4, we present our theoretical radio
afterglow light curves and evaluate the detectability of FAST.
Our conclusions and discussion are given in Section 5.

\section{Sample}

As of Dec 2013, the number of observed radio afterglows is about 95, which is only $\sim30$ \%
of all the GRBs with afterglows being observed. In practice, the observed radio emission should
be composed of the GRB afterglow component and the contribution from the host galaxy.
For GRB 980703, Berger, Kullarni \& Frail (2001b) found that there is a constant component among
the observed radio emission, which was interpreted as the first example of host contribution
to the afterglow (see also Frail et al. 2003). They argued that radio and submillimetre
observations of the GRB host galaxies will be very useful for studying the obscured star formation
rate and the properties of starbursts at high redshifts. Berger et al. (2003a) pointed out that
about $20$ \% of GRB host galaxies are ultra--luminous ($L > 10^{12} L_{\sun}$) and they can be
utilized to probe a representative population of star--forming galaxies.

For bursts with detected radio afterglows, we collect the observational data of
host galaxies together with their peak fluxes of radio afterglows at $\nu=1.4\sim9.0$ GHz.
We get 50 GRBs in total, and the data are listed in Table 1.
In our sample, there are 10 GRBs associated with supernovae (SNe).
According to their isotropic energies, we classify the GRBs into three types, namely low-luminosity, standard
and high-luminosity GRBs. The boundary energies of our classification are $10^{51}$ erg and $10^{53}$ erg,
respectively. 6 GRBs without known redshifts are not included in the above three sub-samples,
but are treated as an independent sub-sample listed in Table 1.

In Column 1 of Table 1, the GRB names are given. The isotropic energies $E_{iso}$, the $T_{90}$ durations
in the observer's frame and the redshifts $z$ are listed in Columns 2, 3 and 4, respectively.
Column 5 is the observational frequency. Column 6 tabulates the contribution from the host galaxies.
In this column, 10 radio fluxes for 7 host galaxies have been directly reported
in the literature. Other host fluxes are based on the assumption that the host flux density is constant
at all stages of the afterglow so that the host flux dominates at late times.
We determine the host contribution only for those GRBs which were monitored in radio bands
for a long period and whose radio light curves obviously become flat at the final stages.

Columns 7 and 8 tabulate the observed peak flux densities and the corresponding time of the peak.
Most of the peak flux densities and the peak time are taken from Chandra \& Frail (2012) which provided a sample with many events.
They are determined by fitting the observed light curves. Meanwhile, for 13 GRBs whose peak flux
densities are marked with daggers, the peak flux densities are simply the brightest measurements.
This may cause the peak flux density to be somewhat under-estimated, but not by enough to affect our results.
The references of the host fluxes and the observed peak fluxes are listed in corresponding order in Column 9.
The data observed by Frail \& Chandra have been kindly offered to us by them.
It should be noted that for some GRBs, other researchers may give different reports of host fluxes in the literatures.
For example, Berger et al. (2003b) and Soderberg et al. (2007) also reported flux densities for GRBs 020405 and 050416A,
which are slightly different from the data in Table 1. But these differences are generally not large enough to affect
our results.

To consider the contribution from the host galaxies, we introduce a new parameter --
the Ratio of the radio fluxes (RRF) of the host to the peak emission.
The RRF is defined as
\begin{equation}
RRF\equiv\frac{F_{host}}{F_{o,peak}}=\frac{F_{host}}{F_{host}+F_{b,peak}}.
\end{equation}
The observed peak flux in Table 1, $F_{o,peak}$, is the sum of $F_{host}$ and $F_{b,peak}$,
where $F_{host}$ and $F_{b,peak}$ stand for the host flux density and the peak flux density
of the pure radio afterglow component, respectively. The RRF values are given in Column 10
of Table 1. In this treatment, we also assume that the host flux density keeps constant in all
the post-burst stages, as Berger, Kullarni \& Frail (2001b) once did.

Using the RRF parameter, we can analyze the flux-flux correlation between the host galaxies and
the GRB afterglows. It is interesting to note that our RRF analysis is completely independent
of the distance (redshift) measurement.

\section{Ratio of Radio Flux Densities}

\begin{figure*}
\centering
  \includegraphics[width=7in]{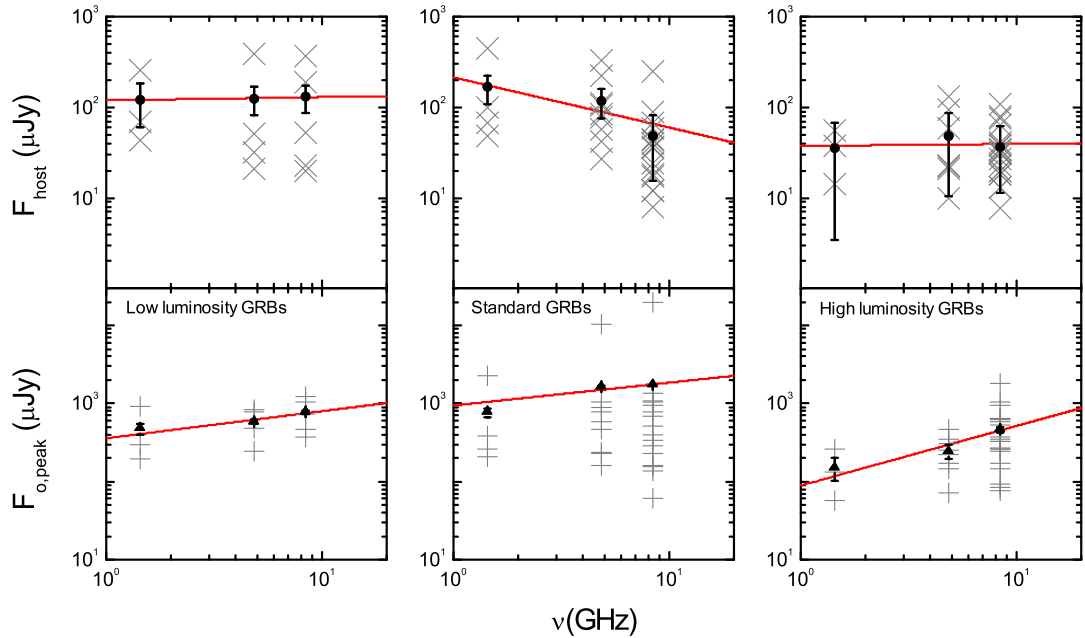}
  \caption{Observed host flux densities (upper panels) and peak flux densities (lower panels)
          as a function of the radio frequencies for low-luminosity (left), standard (middle),
          and high-luminosity GRBs (right), and our best fit to the observations.
          Filled circles with error bars are the mean values of the host flux densities at the corresponding frequency.
          Filled triangles with error bars are the mean values of the observed peak flux densities.
          The solid line in each panel represents our best fit to the observations.
          For the observational data and their references, please see Table 1 for details.}\label{}
\end{figure*}

\begin{figure*}
  \centering
  \includegraphics[width=5.2in]{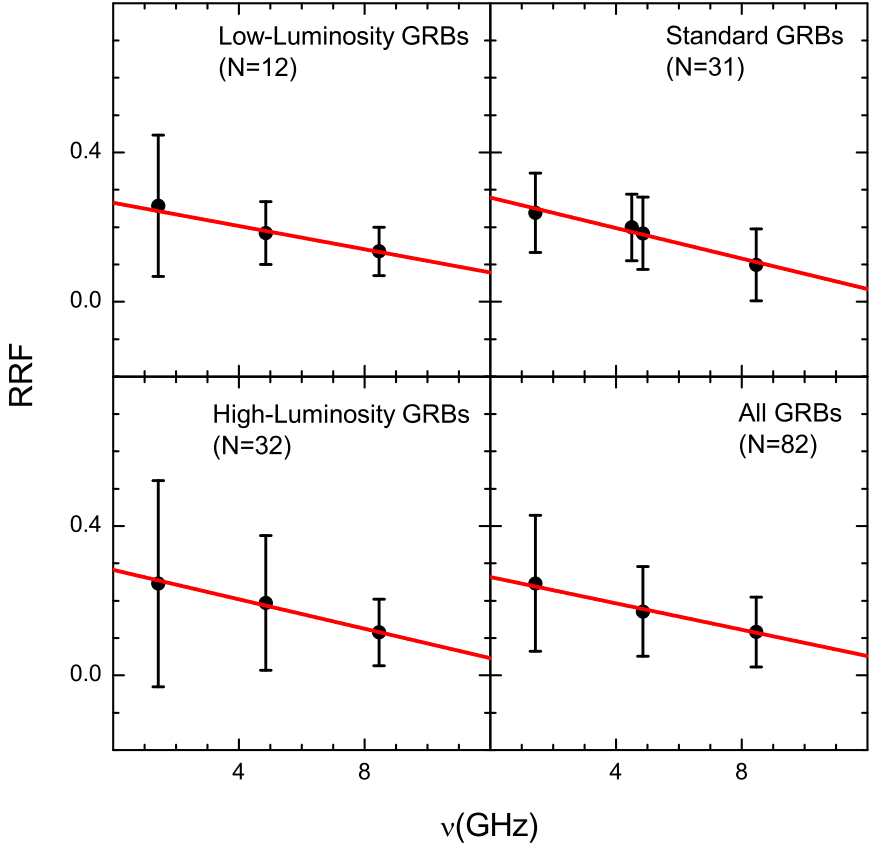}
  \caption{The best linear fits to RRF versus $\nu$ for the low-luminosity, standard,
           high-luminosity GRBs and all GRBs as listed in Table 1. The observed frequencies $\nu$
           are the three corresponding frequencies, in addition, one averaged data point at a
           frequency of 4.5 GHz is also included in the second panel. N in each panel is the
           number of original data points used to derive the mean observational values (solid circles).}\label{}
\end{figure*}

\begin{table*}
  \renewcommand \thetable {2}
  \centering
  \begin{minipage}{129mm}
  \caption{Best fit parameters for the linear RRF --- $\nu$ correlation.}
  \begin{threeparttable}
  \begin{tabular}{@{}lcccc@{}}
  \hline
   GRB types & \multicolumn{4}{|c|}{Fitting Results} \\

  \hline \hline

             & a & b & correlation coefficient & P-value  \\
  \hline
  Low Luminosity GRBs  & $0.27\pm0.02$  & $-0.016\pm0.002$  & 0.96  & 0.09  \\
  Standard GRBs        & $0.28\pm0.02$  & $-0.020\pm0.003$  & 0.95  & 0.02  \\
  High Luminosity GRBs & $0.28\pm0.01$  & $-0.019\pm0.002$  & 0.98  & 0.06  \\
  All GRBs             & $0.26\pm0.01$  & $-0.018\pm0.002$  & 0.98  & 0.07  \\
  \hline
  \end{tabular}
  \end{threeparttable}
  \end{minipage}
\end{table*}

Considering the fact that most current GRB radio afterglows are observed at 1.43, 4.86 and 8.46 GHz,
we use all data in these three bands except those without known redshifts in Table 1.
In Figure 1, we plot the observed host fluxes and peak fluxes versus frequency for different types of GRBs.
The data points are relatively scattered. We then calculate the mean values for the
two fluxes at the three frequencies by assuming an equal weight for each data point in logarithmic scale.
Using the mean values of the host flux densities and the peak flux densities, we fit the correlations
between the fluxes and the observational frequencies for low-luminosity, standard and high-luminosity GRBs, respectively.
The results are also plotted in Figure 1. Generally, the spectra of radio afterglows and host galaxies are
usually considered to be a power-law function of frequency, so we select the power-law fit.
In the three upper panels, the correlations are
$F_{host}\propto\nu^{+0.04\pm0.02}$, $F_{host}\propto\nu^{-0.55\pm0.27}$ and
$F_{host}\propto\nu^{+0.02\pm0.16}$ for low-luminosity, standard and high-luminosity GRBs,
with the correlation coefficients and P-values (rejection possibilities)
being (0.47, 0.0003), (0.60, 0.004) and (-0.95, 0.003) correspondingly.
The three lower panels show that the observed peak fluxes are also correlated with the frequency as
$F_{o,peak}\propto\nu^{+0.34\pm0.13}$, $F_{o,peak}\propto\nu^{+0.29\pm0.18}$
and $F_{o,peak}\propto\nu^{+0.76\pm0.24}$ for the three types of GRBs, with the correlation
coefficients and P-values being (0.80, 0.001), (0.61, 0.001) and (0.90, 0.002), respectively.

Theoretically, the radio flux density of a GRB afterglow can be simplified as a power-law
function of time and frequency, say, $F\propto t^{\alpha}\nu^{\beta}$ (e.g. Sari, Piran \& Narayan 1998;
Wu et al. 2005; Paper I), where $\alpha$ and $\beta$ are the temporal and spectral indices,
respectively. Generally, the number density of the surrounding medium is assumed to depend on
the shock radius as $n\propto R^{-k}$. Liang et al. (2013) and Yi, Wu \& Dai (2013) argued that
the $k$ value should be in the range of $0.4-1.4$ for many GRBs, with a typical value of $k\sim1$.
According to Wu et al. (2005), in the case of homogenous ISM with $k=0$, $F_{b,peak}\propto\nu^{0}$.
But in the case of a typical stellar wind environment with $k=2$, $F_{b,peak}\propto\nu^{1/3}$,
assuming the radio peak is caused when the synchrotron self-absorption frequency, $\nu_{a}$, crosses the observational band.
That is to say, the range of the power-law index for the $F_{b,peak}-\nu$ relation would be $0-1/3$.
Our fitting results are consistent with this range.

In our study, we average the original RRF values of the individual GRBs at a frequency to
get a mean RRF value by assuming an equal weight for each observed data point.
These mean values are then used in our final fitting process.

For each kind of GRBs, we originally adopt two fitting functions, the linear function and the power-law function,
to investigate the correlation between RRF and the observational frequency $\nu$.
The detailed fitting functions are
\begin{equation}
RRF=g(\nu)=\left\{
              \begin{array}{ll}
                a+b\nu, & \hbox{Linear Case}, \\
                          &   \\
                A\nu^{B}, & \hbox{Power-law Case}.
              \end{array}
           \right.
\end{equation}
The power-law case is proved to be worse than the linear one and will have been neglected subsequently.
In Figure 2, we present our best fits in the linear case for the three different types of GRBs
and all the GRB sample as listed in Table 1.
The derived fitting parameters ($a$ and $b$), the correlation coefficients and the P-values
are listed in Table 2. The correlation coefficients are generally high, and the P-values are small, indicating
that the correlation is real and is not a phenomenon by chance.
For the different sub-samples of GRBs in Table 1, we also use the Jackknife resampling method to test the linear model
and find that the linear model is reasonable in a statistical sense.
In Figure 3, we plot the best linear fit lines for all the subsamples. We can see clearly that the four lines
differ from each other only slightly.

In fact, according to the definition, RRF should be within the range of 0 -- 1.
An RRF near 0 indicates that the afterglow component dominates the observed flux,
while an RRF close to 1 indicates that the host flux is much larger than $F_{b,peak}$,
which generally happens in low frequency ranges.

\begin{figure}
 \centering
  \includegraphics[width=9cm]{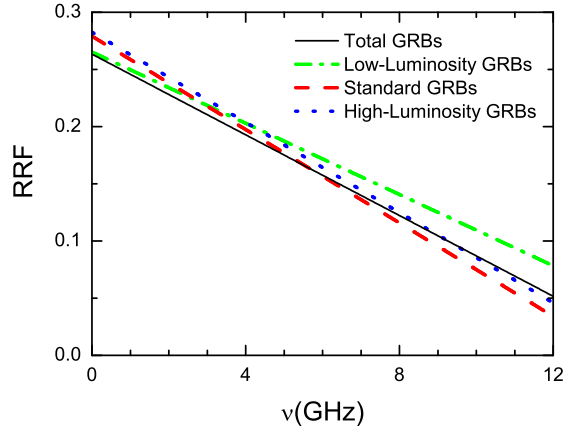}
  \caption{Comparison of the best linear fits of the RRF-$\nu$ correlation
           for different kinds of bursts. The dash-dotted, dashed, dotted and solid lines represent
           the fit to low-luminosity, standard and high-luminosity GRBs and all GRBs, respectively.}\label{}
\end{figure}

\begin{table*}
  \renewcommand \thetable {3}
  \centering
  \begin{minipage}{140mm}
  \caption{Typical physical parameter values used in theoretically modeling of GRB afterglows.}
  \begin{threeparttable}
  \begin{tabular}{@{}lclccccccc@{}}
  \hline
   Source &  $E_{iso}$  & $\gamma_0$ & $\theta_{0}$ & n & p & $\xi^2_{B}$ & $\xi_{e}$  & $\epsilon$ & $\theta_{obs}$ \\

    & (erg) &  & (rad) & ($cm^{-3}$) & & &  & & (rad) \\

  \hline
  Standard GRBs\tnote{a}\,\,...............
                                   & $1.0\times10^{52}$ & $300$ & 0.10 & 1.0 & 2.5 & $1.0\times10^{-3}$ & 0.10 & 0 & 0  \\
  Failed GRBs\tnote{a}\,\,....................
                                   & $1.0\times10^{52}$ & $30$  & 0.10 & 1.0 & 2.5 & $1.0\times10^{-3}$ & 0.10 & 0 & 0  \\
  High-Lominosity GRBs\tnote{a}\,\,...
                                   & $1.0\times10^{54}$ & $300$ & 0.10 & 1.0 & 2.5 & $1.0\times10^{-3}$ & 0.10 & 0 & 0  \\
  Low-Luminosity GRBs\tnote{a}\,\,....
                                   & $1.0\times10^{49}$ & $300$ & 0.10 & 1.0 & 2.5 & $1.0\times10^{-3}$ & 0.10 & 0 & 0  \\

  \hline
  \end{tabular}
  \begin{tablenotes}
       \footnotesize
       \item[a] Refer to Paper I.
  \end{tablenotes}
  \end{threeparttable}
  \end{minipage}
\end{table*}

Using Eqs. (1) and (2), for the linear case, one can easily derive
\begin{equation}
   F_{host}=g(\nu) F_{o,peak}=(a+b \nu)F_{o,peak}.
\end{equation}
Eq. (3) can be used to calculate the host fluxes of those GRBs with radio afterglows being detected
but no hosts being identified. We can also get
\begin{equation}
   F_{host}=\frac{g(\nu)}{1-g(\nu)}F_{b,peak}
           =\frac{a+b \nu}{1-(a+b \nu)}F_{b,peak}.
\end{equation}
Eq. (4) can also potentially be used to predict the flux density of the peak afterglow
or the host at particular frequencies when no direct observational data are available.

This RRF-predicted host fluxes can help us to subtract the potential radio background emission from a GRB source
so that the pure afterglow component can be identified, especially at lower frequencies.
The RRF-$\nu$ correlation can also explain why the low-frequency radio afterglows are usually difficult to be detected,
in contrast with the high-frequency case. According to the RRF-$\nu$ relations, radio fluxes are dominated by
the host component at lower frequencies, as shown in Figures 1-3.

\section{Theoretical Radio Afterglows}

According to Eqs. (2), (3) and (4), the host flux density depends on the observing frequency.
Thus, we can use the RRF at a certain frequency to calculate the host flux density
if the peak flux density of the radio afterglow is available.
Here we use the theoretical peak flux density calculated from the fireball-shock model
to estimate the host flux density. Taking into account the contribution from the
host galaxy, we will then study the detectability of radio afterglows by FAST.
As Zhang et al. (2015) did in Paper I, we also calculate the radio afterglow light curves
for four kinds of GRBs, i.e. standard, failed, high luminosity, and low luminosity GRBs.

\subsection[]{Dynamics}

Here, we adopt the generic dynamical equations for beamed GRB outflows (e.g. Huang et al. 1998, 1999a, 1999b, 2000a, 2000b)
to calculate the radio afterglows of GRBs. The formulae of the dynamical model were widely applied in many studies
such as the overall afterglow modeling (Cheng \& Wang 2003; Wu et al. 2004; Kong et al. 2009),
the rebrightening at multi-wavelengths (Dai et al. 2005; Xu \& Huang 2010; Kong et al. 2010; Yu \& Huang 2013; Hou et al. 2014),
and the beaming effect (Huang, Dai \& Lu 2000c; Wei \& Lu 2002; Huang et al. 2003), etc.
These equations are valid in both the ultra-relativistic and non-relativistic stages, thus they are especially convenient
for calculating radio afterglows which usually last for a very long period and may involve the Newtonian regime.
In our calculations, the effects of electron cooling, lateral expansion and equal arrival time surfaces are included.

The parameters assumed in our calculations are listed in Table 3,
where
 $E_{iso}$ is the initial isotropic energy, $\gamma_0$ is the initial bulk Lorentz factor,
 $\theta_{0}$ is the initial half-opening angle of the jet,
 $\theta_{obs}$ is the angle between the axis of the jet and the line of sight,
 $n$ is the number density of surrounding ISM,
 $p$ is the electron distribution index,
 $\xi_{e}$ and $\xi^2_{B}$are respectively the energy
 fractions of electrons and magnetic field with respect to the total energy.
The radiative efficiency $\epsilon$ equals 1 in the highly radiative case and equals 0 in the adiabatic case.
The initial Lorentz factor and isotropic energy are evaluated differently for the four kinds of GRBs,
while the following common parameters are assumed to be universal, namely
$n=1 cm^{-3}, p=2.5, \xi_{e}=0.1, \xi^2_{B}=0.001, \theta=0.1$ and $\theta_{obs}=0$.
Meanwhile, we take $\epsilon=0$ since we are mainly considering the late-time afterglows,
which should be in the adiabatic regime.

\subsection[]{Numerical Results}

\begin{table}
  \renewcommand \thetable {4}
  \centering
  \caption{Parameters of the 9 sets of FAST's receivers.}
  \begin{threeparttable}
  \begin{tabular}{@{}cccc@{}}
  \hline
  No. & Bands\tnote{a} & $\nu_{c}$\tnote{a} & flux density limit\tnote{a} \\
      & (GHz)   &  (GHz)      &  ($\mu Jy$)  \\
  \hline
  1 & 0.07 - 0.14   & 0.10  & 843.5  \\
  2 & 0.14 - 0.28   & 0.20  & 238.0  \\
  3 & 0.320 - 0.334 & 0.328 & 376.5  \\
  4 & 0.28 - 0.56   & 0.40  & 63.5   \\
  5 & 0.55 - 0.64   & 0.60  & 44.5   \\
  6 & 0.56 - 1.02   & 0.80  & 18.0   \\
  7 & 1.23 - 1.53   & 1.38  & 9.5    \\
  8 & 1.15 - 1.72   & 1.45  & 7.0    \\
  9 & 2.00 - 3.00   & 2.50  & 5.5    \\
  \hline
  \end{tabular}
  \begin{tablenotes}
       \footnotesize
       \item[a] These data are taken from Nan et al. (2011) and Paper I.
                The flux density limit is the 5$\sigma$ detection limit of FAST
                for a 30-minute integration time.
  \end{tablenotes}
  \end{threeparttable}
\end{table}

In order to make a comparative analysis with Paper I, we re-calculate the radio afterglow
light curves for four types of GRBs according to FAST's nine passbands,
assuming a typical redshift of $z=1.0$. As in Paper I, we also adopt the limiting flux densities
under an integral time of 30 minutes as the limiting sensitivity for FAST.
The difference is that we now include the contribution from the host, whose flux is estimated
from the linear RRF-$\nu$ correlation. Our numerical results are plot in Figure 4.
We see that the host flux densities estimated from Eqs. (2) and (4) are often significantly larger than
the afterglow component at early and late times, which makes the total light curves
much flatter than that in Paper I which did not consider the host galaxy contribution.
This would make the radio afterglows more difficult to be identified due to the influence of the
relatively strong radio background.

\begin{figure*}
  \centering
  \includegraphics[width=6in]{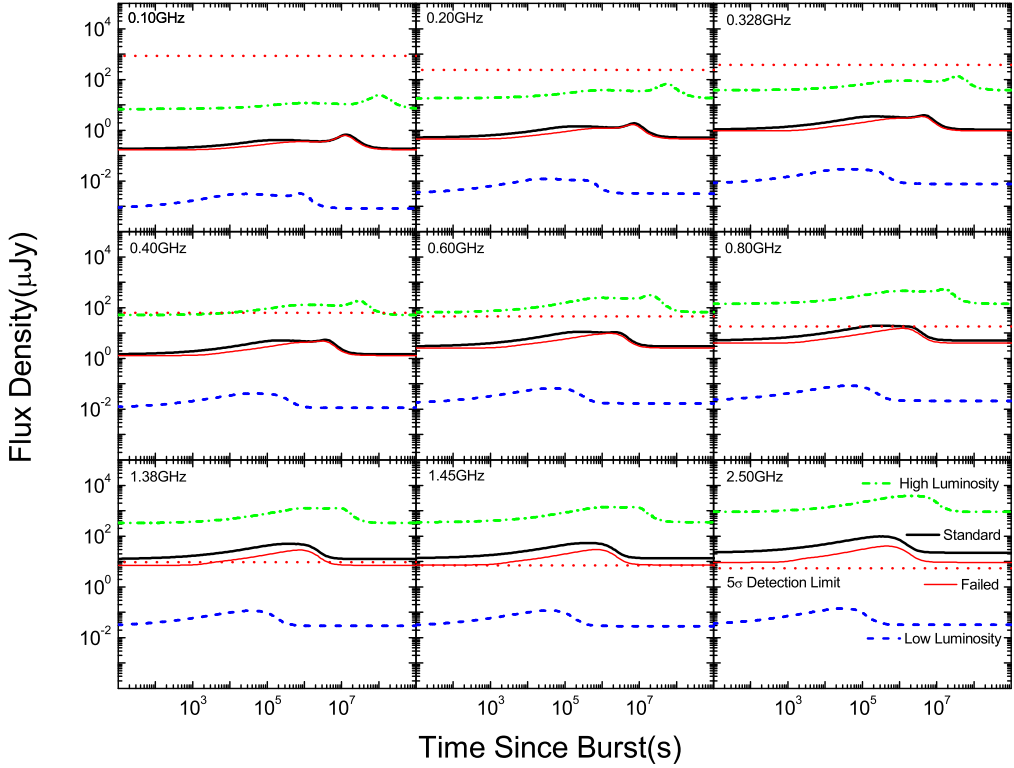}\\
  \caption{Predicted radio light curves within FAST's nine energy passbands for four kinds of GRBs
           lying at $z=1.0$ for the linear RRF-$\nu$ case.
           The thin solid, dash-dotted, dashed and thick solid lines correspond to failed, high-luminosity,
           low-luminosity and standard GRBs, respectively. The horizontal dotted lines represent the
           5$\sigma$ limiting sensivities of FAST for a 30-minute integration time (Paper I).
           Note that the host flux densities predicted by the linear RRF-$\nu$ correlation have been added
           in these light curves.}\label{}
\end{figure*}

\begin{figure}
 \centering
  \includegraphics[width=9.2cm]{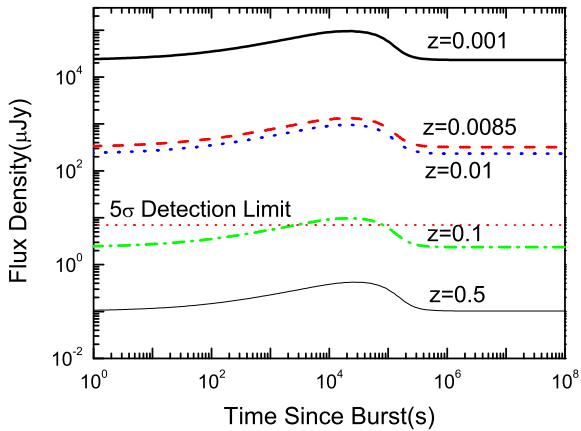}
  \caption{Predicted radio light curves of low-luminosity GRBs at 1.45GHz for the linear RRF-$\nu$ case.
           The redshifts for the thick solid, dashed, dotted, dash-dotted and thin solid lines are $z=0.001,
           0.0085$ (the redshift of GRB 980425), $0.01, 0.1$ and $0.5$, respectively. The horizontal dotted line
           represents 5$\sigma$ limiting sensivity of FAST for a 30-minute integration time (Paper I).
           Note that the host flux densities predicted by the linear RRF-$\nu$ correlation have been added
           in these light curves.}\label{}
\end{figure}

According to Figure 4, standard GRBs could be observed by FAST at frequencies $\nu > 0.80$ GHz,
though at $\nu = 0.80$ GHz, the predicted peak flux density is 19.6 $\mu$Jy, and the $5\sigma$
detection limit of FAST for a 30-minute integration time is 18.0 $\mu$Jy.
Note that high luminosity GRBs have the brightest radio afterglows so that they can be detected
easily in most bands with $\nu>0.40$ GHz.
At $\nu=0.40$ GHz, FAST can detect high-luminosity GRBs from $\sim0.1$ day to $\sim1270$ days.

An interesting phenomenon displayed in Figure 4 is that the radio afterglows of standard and
failed bursts are very different at frequency $\nu \geq 1.38$ GHz. The reason is that
the initial bulk Lorentz factor of failed GRBs is about ten times smaller than that of standard ones.
A lower Lorentz factor then leads to a lower flux density at early times as well as a lower peak flux density.
However, the late time afterglow depends mainly on the intrinsic energy of the jet, so the light curves
differ from each other only slightly at late stages, especially at low frequencies (Wu et al. 2004).
The radio emission can be detected from $\sim4000$ s to $\sim 58$ days at $\nu=1.38$ GHz.
At $\nu\geq1.45$ GHz, FAST can potentially detect very early radio emission.

Radio afterglows of low-luminosity bursts are the weakest ones in each panel of Figure 4.
For example, the strongest peak flux is only 0.14 $\mu$Jy at $\nu=2.50$ GHz.
According to our calculations, FAST can hardly detect any radio afterglows of low-luminosity GRBs
at the redshift of $z=1$. However, many low-luminosity GRBs are likely to happen very near to us,
then they can also be observed by FAST. A good example is the low-luminosity GRB 980425 ($z=0.0085$).
At such a small distance, its radio brightness is very high. As shown in Figure 5,
radio afterglows of low-luminosity GRBs can be detected by FAST at a redshift of $z<0.1$.
Fan, Piran \& Xu (2006) argued that such under-luminous GRBs with less kinetic energy might be
very common. We expect that much more low-luminosity GRBs would be detected by FAST in the future.

\section{discussion and conclusions}

In this study, we investigate the connection between host fluxes and peak afterglow fluxes in
radio bands statistically.
The observed GRBs are classified into three types, i.e. low-luminosity, standard and high-luminosity GRBs.
It is found that there is an anti-correlation between the RRF and the observational frequency.
At a higher frequency, the corresponding RRF is smaller. This could be due to more significant
self-absorption at longer wavelengths (Rybicki \& Lightman 1979).
Based on this correlation, the host flux densities at different radio frequencies can be estimated.
This RRF prediction is especially helpful for those GRBs whose radio afterglow data are very limited.
Meanwhile, we also re-considered the capability of detecting GRBs with FAST.
FAST have 9 energy channels in its first phase, ranging from 0.10 GHz to 2.50 GHz (Nan et al. 2011).
It will mainly operate in a relatively low frequency range.
Our results show that at a typical redshift of $z=1.0$,
the radio emission of standard GRBs can be well detected by FAST at $\nu\,\geq 1.38$ GHz.
Although FAST can hardly detect radio afterglows from low-luminosity GRBs located at this redshift,
we expect that a large population of low-luminosity GRBs could still be observable since many such
events may actually happen much nearer to us.

The host galaxies of long GRBs can help to reveal the environments
and populations of the unusual GRB progenitors (Levesque 2014).
Berger (2014) argued that the properties of short GRBs' hosts will hint on the
redshift distribution and shed light on the progenitors' age distribution.
Berger, Kullarni \& Frail (2001b) proposed that if more host galaxies are detected and
studied in detail in the radio and submillimeter/FIR wavelengths, we will be able to address a large
number of issues pertaining not only to the bursts themselves but also to the characteristics
of galaxies at high redshifts. This will lead to a thorough understanding of GRB trigger mechanisms
and the host environments.
Unfortunately, the detection rate of afterglows in radio bands is only $\sim30$\%,
which is significantly lower than that of optical and X-ray afterglows.

In Paper I, the peak flux density versus the peak frequency for the above mentioned four types
of GRBs at different redshifts were plot theoretically.  It was shown that the radio flux density $F$ is a
power-law function of $F\propto\nu^{2}$ (e.g. Sari, Piran \& Narayan 1998; Wu et al. 2005)
if the observing frequency $\nu$ is below the synchrotron self-absorption frequency $\nu_{ssa}$.
The dependence of the radio afterglow light curves on various parameters has been investigated by some authors,
such as Chandra \& Frail (2012). They found that the radio afterglow is relative bright when the ISM density
is between $n=1$ and $10\,cm^{-3}$. This may explain why some GRBs bright in X-ray/optical bands
are dim in radio bands. Also, the radio brightness strongly depends on the intrinsic energy of the outflow.
So, there is evidently a close relationship between the detectability of the radio afterglow
and the intrinsic physical parameters of the GRB (Zhang et al. 2015).

\section*{Acknowledgments}

We thank the anonymous referee for valuable comments and suggestions that lead to an overall improvement of this study.
We acknowledge D. A. Frail and P. Chandra for kindly offering their invaluable radio afterglow data to us.
We are thankful to B. Zhang, B.B. Zhang and Y. Z. Fan for helpful discussions. This work is partly supported
by the National Basic Research Program of China (973 Program, Grant No. 2014CB845800) and the National Natural
Science Foundation of China (Grant Nos. 11263002, 11473012 and 11322328). XFW and DL acknowledges support by
the Strategic Priority Research Program ``The Emergence of Cosmological Structures'' (Grant No. XDB09000000) of the
Chinese Academy of Sciences. SWK acknowledges support by China Postdoctoral science foundation under grant 2012M520382.
HYC was supported by BK21 Plus of the National Research Foundation of Korea and a National Research Foundation of
Korea Grant funded by the Korean government (NRF-2013K2A2A2000525).

\appendix
\label{lastpage}

\end{document}